\documentclass[12pt,epsfig]{article} 
\usepackage{epsfig}
\begin{document}

\begin{center}
ANALISYS   OF THE REACTION $np \rightarrow np \pi^+ \pi^-$ FROM THE POINT OF VIEW
 OF OPER-MODEL 
\\
\vspace{0.3cm}
A.P. Jerusalimov
\\
JINR, Dubna, Moscow region, 141980, Russia 
\end{center}

\vspace{0.3cm}
\begin{abstract}
  The reaction $np \rightarrow np \pi^+ \pi^-$ was studied at the various
momenta of incident neutrons. It was shown that the characteristics of the
reaction at the momenta above 3 GeV/c could be described by the model of reggeized
$\pi$~exchange (OPER). At the momenta below 3 GeV/c, it was necessary to use 
additionally the mechanism of one baryon exchange (OBE).   
\end{abstract}

\vspace{1.0cm}
\section{Introduction: study of inelastic np interactions  
                  at  accelerator facility of  LHEP JINR}

  The data about inelastic $np$ interactions were obtained due to irradiation
of 1m hydrogen bubble chamber (4$\pi$ geometry) by quasimonochromatic neutron 
beam ($\delta P < 2.5\%$) at the following incident momenta:\\ 
$P_0$=1.25, 1.43, 1.73, 2.23, 3.10, 3.83, 4.10 and 5.20 GeV/c

 The unique of fullness and precision data are obtained~\cite{np_inel}. 
It permits to carry out the detailed study of inelastic $np$ interactions in a
wide region of energies. 

\begin{figure}[h]
\includegraphics[width=1.0\textwidth]{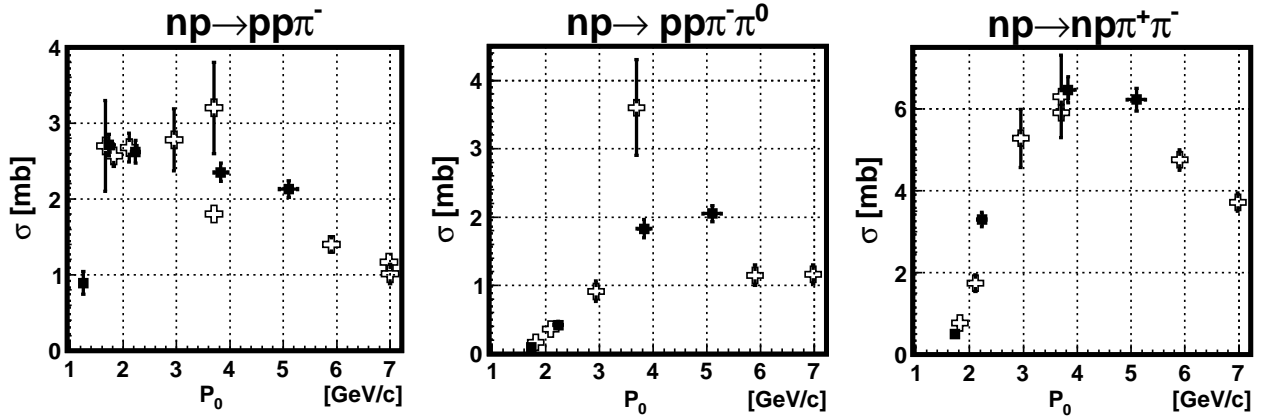}
\caption{Cross-sections of some inelastic np interactions
         (black squares - our data)}
\label{Fig1}
\end{figure}

\newpage
\section{The reaction $np \rightarrow np \pi^+ \pi^-$ at $P_0$ $>$ 3 GeV/c}

  This reaction is characterized by:\\
\hspace*{1.5cm} - plentiful  production  of  the $\Delta$-resonance (see Fig.2),
\\
\begin{figure}[h]
\includegraphics[width=1.05\textwidth]{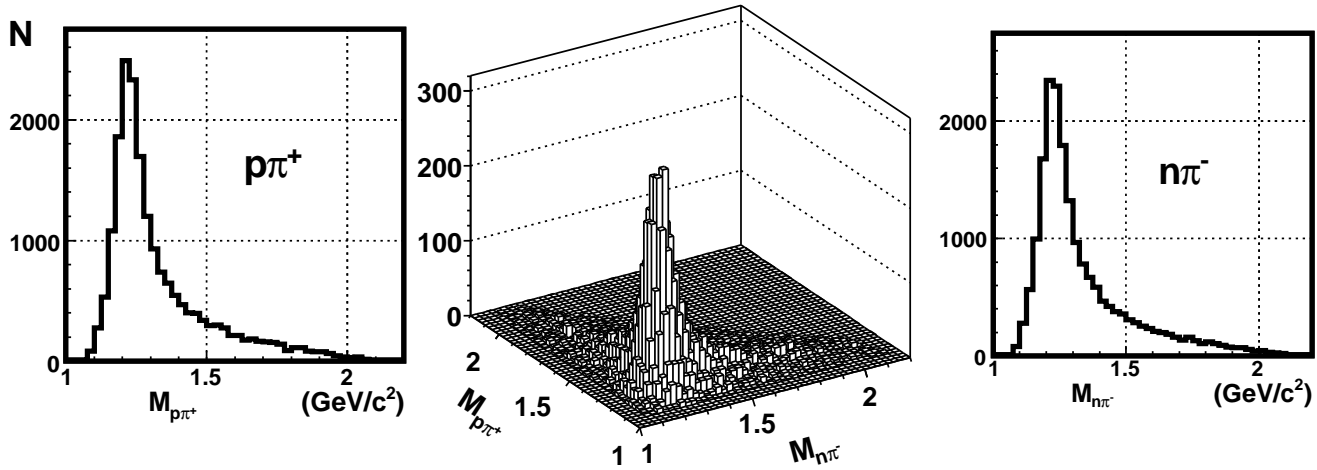}
\caption{The distributions of $M_{p\pi^+}$ and $M_{n\pi^-}$ from the reaction
  $np \rightarrow np \pi^+ \pi^-$ at $P_0$ = 3 GeV/c}
\label{Fig2}
\end{figure}
\hspace*{1.4cm} - large peripherality of the secondary nucleons.

Various modifications of the one pion exchange models (OPE) are used 
to describe the experimental data of the inelastic $NN$, $N\bar N$ и 
$\pi N$-interactions. At that parameters of these models are different 
for various processes and even for concrete reactions at various energies.
Various models differ also in respect of the reggeization of $\pi$-meson:
at times an exchange by elementary $\pi$-meson is used~\cite{Wolf} at other 
times - by reggeized $\pi$-meson~\cite{Berger}.

  The models of Regge pole exchange\cite{Nikitin,Collins} are based on the 
method of  complex momentaand consider  an exchange in t-channel by a virtual
state R that has quantum numbersof particle (resonances) with variable spin and
is on some trajectory $\alpha_R(t)$ named Regge trajectory. According to this 
model, the amplitude of  binary and quasi-bynary  processes such as 
$a+b \rightarrow c+d$ 
\begin{figure}[h]
\hspace{3.0cm}
\includegraphics[width=0.5\textwidth]{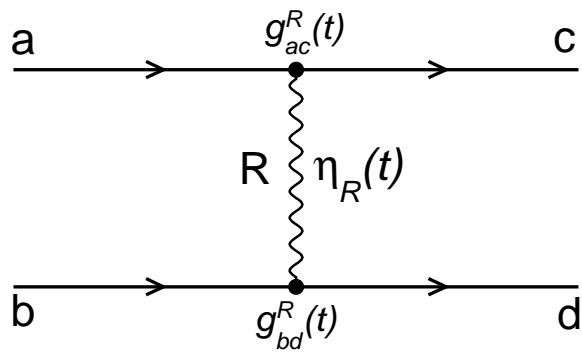}
\caption{Diagram of the process $a+b \rightarrow c+d$}
\label{Fig3}
\end{figure}

is written in the following form [3]:
\begin{equation}
\label{ReggeEx}
  T_R(s,t)=i8 \pi s_0 \; g_{ac}^R(t) \; \eta_R(t) \left( \frac{s s_0}
  {m_c^2 m_d^2} \right)^{\alpha_R(t)} \!\! g_{bd}^R(t)
\end{equation}

where $s_0$ - energy scale factor,\\
\hspace*{1.7cm} $g_{ac}^R(t)$ и $g_{bd}^R(t)$ - vertex functions,\\
\hspace*{1.7cm} $\eta_R(t)$ - signature factor, that is determined in the 
following form:
$$
 \eta_R(t) = - \frac{\sigma + \exp(-i \pi \alpha_R(t))}{\sin{[\pi \alpha_R(t)]}}
$$
Signature $\sigma$ is the quantum number characterizing particles (resonances) 
and correspondingly Regge pole trajectory passing through them. It is determined
by a parity of the particle (resonance):\\
\hspace*{1.1cm}$ \sigma = (-1)^{l_R}$ for integer $l_R$ (bosons),\\
\hspace*{1.1cm}$ \sigma = (-1)^{l_R \pm 1/2}$ for semi-integer $l_R$ (fermions).

The most developed and detailed model of reggeized $\pi$-meson exchange is the 
model suggested in ITEP~\cite{Ponomarev}. The advantages of this model are:\\
\hspace*{1.5cm} - small number of free parameters (3  in our case),\\
\hspace*{1.5cm} - wide region of the described energies (2$\div$200 GeV),\\
\hspace*{1.5cm} - calculated values are automatically normalized to the reaction
 cross-section.\\

 Within the framework of this model the diagrams of the following form give main
contribution into the reaction $np \rightarrow np \pi^+ \pi^-$ :
\begin{figure}[h]
\hspace{3.0cm}
\includegraphics[width=0.5\textwidth]{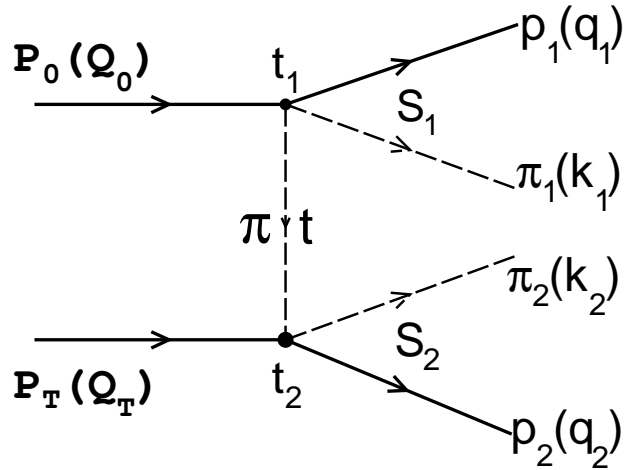}
\caption{Main diagram for the reaction $N N \rightarrow N N \pi \pi$}
\label{Fig4} 
\end{figure}

Let’s introduce the following notations:\\
$s=(Q_1+Q_2)^2$, $s_1=(q_1+k_1)^2$, $s_2=(q_2+k_2)^2$,\\
$t_1=(Q_1-q_1)^2$, $t_2=(Q_2-q_2)^2$, $t=(Q_1-q_1-k_1)^2=(Q_2-q_2-k_2)^2$,\\
In that case the matrix element for this diagrams is written in the following 
form:
\begin{equation}
\label{OPER22}
  M_{22} = \; T_{\pi N \rightarrow \pi N}^{up} \: \frac{F_{22}(s,t,s_1,s_2 ...)}
  {(t-m_{\pi}^2)} \: T_{\pi N \rightarrow \pi N}^{down}
\end{equation}
where $1/(t-m_{\pi}^2)$ - $\pi$-meson propagator that is proportional to 
signature factor $\eta_{\pi}(t)=\exp[-i\pi \alpha_{\pi}(t)/2]/\sin[\pi 
\alpha_{\pi}(t)/2]$ at small $|t|$,\\
\hspace*{1.1cm} $T_{\pi N \rightarrow \pi N}$ - amplitudes of elastic 
$\pi N$-scattering off mass shell\\
\hspace*{1.1cm} $F_{22}(s,t,s_1,s_2 ...)$ - form-factor:
$$
 F_{22}= e^{R_2^2(t-m_{\pi}^2)} \left[ \frac{s}{s_0} \frac{\kappa_1^2}{s_1} 
 \frac{\kappa_2^2}{s_2} \right]^{\alpha_{\pi}(t)},
$$
\hspace*{1.1cm}where $ \kappa_i^2 = k_{i \perp}^2 + m_{\pi}^2 - c(t-m_{\pi}^2)$,\\
Usually energy scale factor is determined as $s_0=1$ GeV$^2$ and Regge 
trajectory of $\pi$-meson as linear one $\alpha_{\pi}(t)=\alpha_{\pi}'(
t-m_{\pi}^2)$.
In that way the used version of OPER-model has 3 free parameters: 
$\alpha_{\pi}'$, $R_2^2$ и $c$.

The following main diagrams correspond to the reaction $np \rightarrow np \pi^+ 
\pi^-$ within the framework of  OPER model:
\begin{figure}[h]
\includegraphics[width=1.0\textwidth]{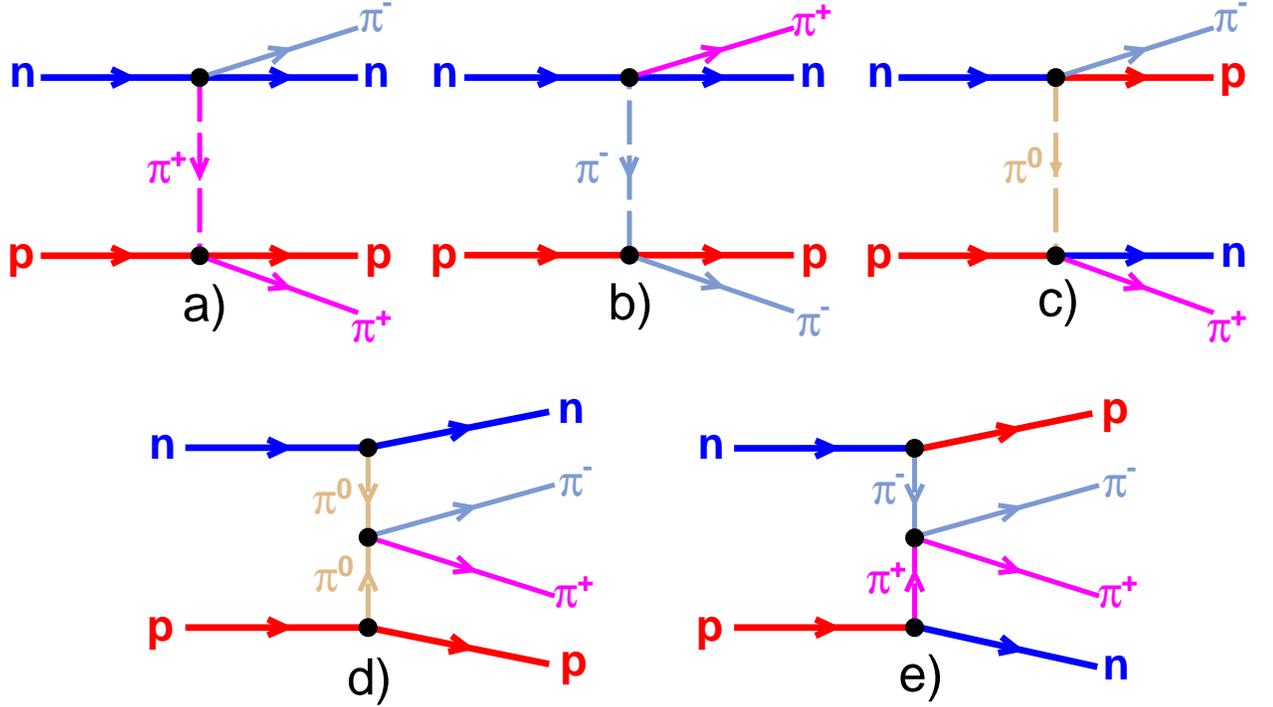}
\caption{OPER diagrams 2$\times$2 for the reaction $np \rightarrow np \pi^+ 
\pi^-$}
\label{Fig5}
\end{figure}

  The study has shown that the contribution of the "hanged" diagrams 
(\ref{Fig5}d) and (\ref{Fig5}e) into the reaction cross-sections at 
$P_0$ $<$ 10 GeV/c is negligible. Interference between diagrams 
(\ref{Fig5}a), (\ref{Fig5}b) and (\ref{Fig5}c) is small and 
do not exceed some $\%$ at $P_0$ $<$ 10 GeV/c.

The use of some specific kinematic cuts similar to used in~\cite{Apeldoorn} 
permits to select the kinematic region of the reaction $np \rightarrow np \pi^+ 
\pi^-$ in which the contribution of the diagram (\ref{Fig5}a) is 
dominating. This approach permits to determine the parameters of the model more 
precisely.

   The slope of the  $\pi$-meson trajectory was taken equal $\alpha_{\pi}'=0.7$ 
(as in~\cite{Ponomarev}). However some modifications of the model were made to 
describe the experimental characteristics of the reaction $n p \rightarrow n p 
\pi^+ \pi^-$.  In particular it was determined that it is better to replace 
parameter $c=0.08$ in expression for $\kappa_i^2$
by $c=\frac{2m_{\pi}}{\sqrt{s}-2m_N}$. Moreover the amplitude of elastic piN 
scattering off mass shell should be written in the form
$$T_{\pi N \rightarrow \pi N}^{off}=\sqrt{\frac{Q(s_i,t_i,t)}{Q(s_i,t_i,
m_{\pi}^2)}} \;\; T_{\pi N \rightarrow \pi N}^{on}$$
  The value of the amplitude of the elastic $\pi N$-scattering on mass shell is 
calculated using the data of partial wave analysis (PWA)~\cite{PWA}. Parameter 
$R_2^2$ was taken equal $3.3$ GeV$^{-2}$.

\begin{figure}[h]
\includegraphics[width=1.0\textwidth]{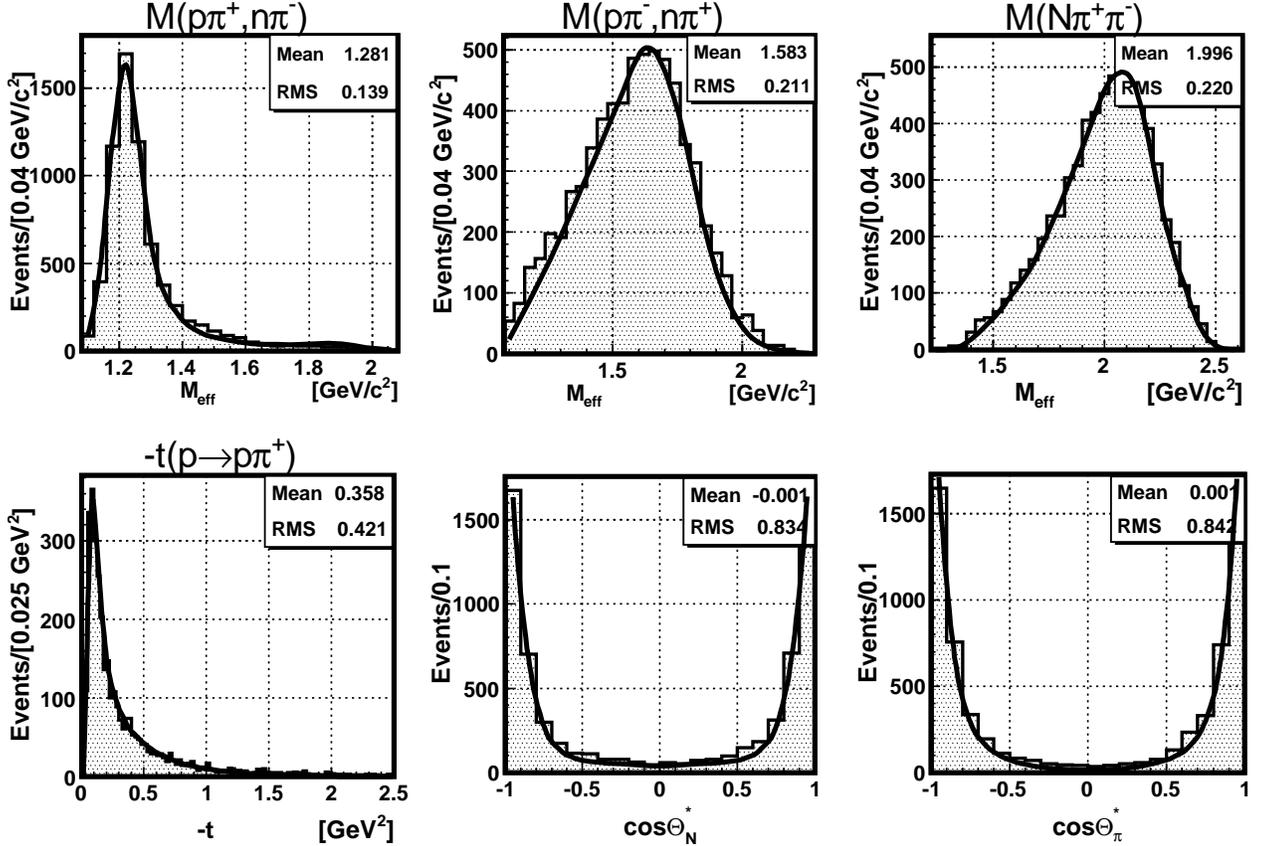}
\caption{Distributions for the reaction $np \rightarrow np \pi^+ \pi^-$ at 
$P_0$=5.20 GeV/c
 obtained due to specific cuts.}
\label{Fig6}
\end{figure}

   The results of the calculations using OPER-model with such set of the 
parameters are shown in Fig.\ref{Fig6} for the selected kinematic region of 
the reaction at $P_0 = 5.2$ GeV/c.

  One can see a good agreement between the experimental distributions and 
theoretical calculations.

  However it is insufficient the diagrams (\ref{Fig5}a), 
(\ref{Fig5}b) and (\ref{Fig5}c) to describe the reaction $n p 
\rightarrow n p \pi^+ \pi^-$ in full kinematic region. It is necessary to take 
into account the diagrams that include the vertex
of inelastic $\pi N \rightarrow \pi \pi N$-scattering shown in 
Fig.\ref{Fig7}:
\begin{figure}[h]
\hspace{1.5cm}
\includegraphics[width=0.80\textwidth]{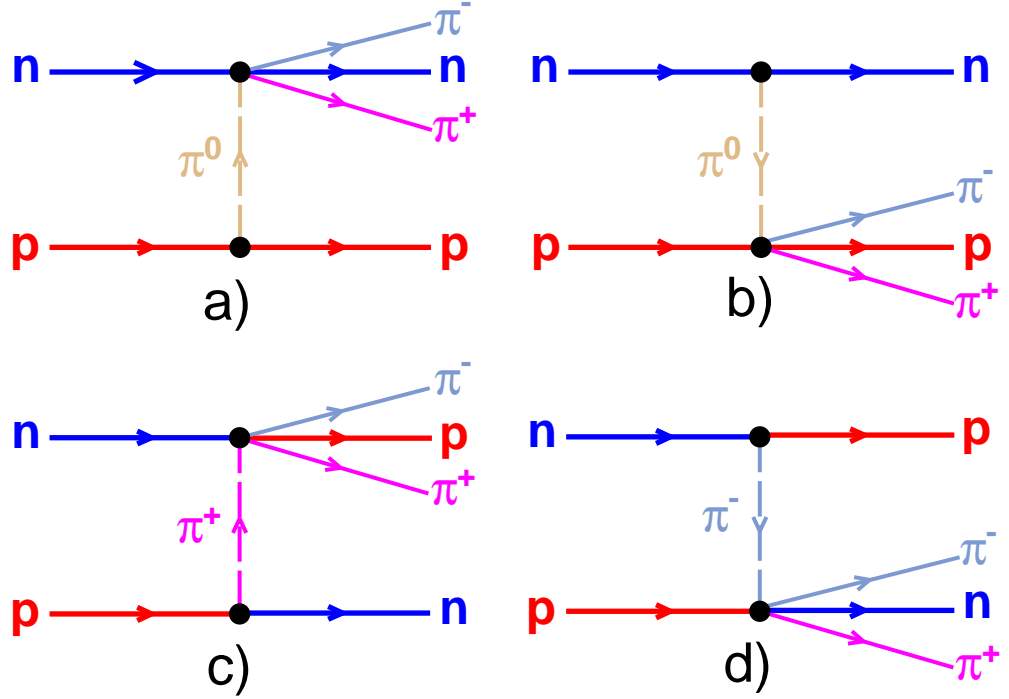}
\caption{OPER diagrams 1$\times$3 for the reaction $np \rightarrow np \pi^+ 
\pi^-$}
\label{Fig7}
\end{figure}

  The matrix element of these diagrams is written in the form similar to the 
diagrams of the reaction $N N \rightarrow N N \pi$ in~\cite{Ponomarev}:
\begin{equation}
\label{OPER13}
M_{13} = G \bar u(q_N)\gamma_5 u(Q_N') \; \frac{F_{13}}{(t - m_{\pi}  
 ^2)} \; T_{\pi N \rightarrow \pi \pi N}
\end{equation}
where  $T_{\pi N \rightarrow \pi \pi N}$ - amplitudes of inelastic 
$\pi N$-scattering off mass shell\\
\hspace*{3.0cm} $T_{\pi N \rightarrow \pi \pi N}^{off}=\sqrt{\frac{Q(s_i,t_i,t)}
{Q(s_i,t_i,m_{\pi}^2)}} \;\; T_{\pi \pi N \rightarrow \pi N}^{on}$;\\
\hspace*{1.2cm} $G \bar u(q_N)\gamma_5 u(Q_N)$ ($N\pi N$)-vertex 
$(G^2/4\pi=14.6)$;\\
\hspace*{1.2cm} $
 F_{13}= e^{R_1^2(t-m_{\pi}^2)} \left[ \frac{s}{s_0} 
\frac{\kappa^2}{s_{N\pi\pi}} \right]^{\alpha_{\pi}(t)}$ - form-factor,\\
\hspace*{1.3cm} $R_1^2$ - free parameter.\\

 The calculation of the amplitudes of $\pi N \rightarrow \pi \pi N$ reactions
are described in~\cite{pipiN}.

  It is significant to mention a detail in the determination of  value 
$\kappa^2$ in the formfactor $F_{13}$. The value $ \kappa^2 = k_{\pi \perp}^2 +
 m_{\pi}^2 - c(t-m_{\pi}^2)$ for the reaction $N N \rightarrow N N \pi$. 
But the parameterization of the reaction $\pi N \rightarrow \pi \pi N$ 
assumes that it is in fact the sum of separate 2-particle channels 
(see Appendix in~\cite{pipiN}):\\
\hspace*{4.5cm} $\pi N \rightarrow N^*(\Delta^*) \rightarrow \Delta \pi$,\\
\hspace*{4.5cm} $\pi N \rightarrow N^*(\Delta^*) \rightarrow N \rho$,\\
\hspace*{4.5cm} $\pi N \rightarrow N^*(\Delta^*) \rightarrow N \epsilon$,\\ 
\hspace*{4.5cm} $\pi N \rightarrow N^*(\Delta^*) \rightarrow N^*_{1440} \pi.$\\
Then in accordance with~\cite{Ponomarev} there are 4 formfactors:\\
\hspace*{1.0cm} $F_{13 \Delta}$ for $\pi N \rightarrow \Delta \pi$ with 
$ \kappa^2 = k_{\pi \perp}^2 + m_{\pi}^2 - c(t-m_{\pi}^2)$ and 
$c=\frac{m_{\pi}}{\sqrt{s}-2m_N}$;\\
\hspace*{1.0cm} $F_{13 \rho}$ for $\pi N \rightarrow N \rho$ with 
$ \kappa^2 = k_{\rho \perp}^2 + m_{\rho}^2 - c(t-m_{\pi}^2)$ and 
$c=\frac{m_{\rho}}{\sqrt{s}-2m_N}$;\\
\hspace*{1.0cm} $F_{13 \epsilon}$ for $\pi N \rightarrow N \epsilon$  with 
$ \kappa^2 = k_{\epsilon \perp}^2 + m_{\epsilon}^2 - c(t-m_{\pi}^2)$ and 
$c=\frac{m_{\epsilon}}{\sqrt{s}-2m_N}$;\\
\hspace*{1.0cm} $F_{13 N^*}$ for $\pi N \rightarrow N^*_{1440} \pi$  with 
$ \kappa^2 = k_{\pi \perp}^2 + m_{\pi}^2 - c(t-m_{\pi}^2)$ and 
$c=\frac{m_{\pi}}{\sqrt{s}-2m_N}$. \\
  This choice of the formfactor provide an explanation for the absence of the 
clear signal of the $\rho$-meson production in the effective masses of 
$\pi\pi$-combinations from $N N \rightarrow N N \pi \pi$ reactions. The 
channels of the production both $\rho$-meson and $\epsilon$-meson are suppressed
in comparison with the channels of the $\Delta$ and $N^*$ production due to a 
considerably larger values of $\kappa^2$ in formfactor $F_{13}$.

  It was shown in~\cite{Diff} that the processes of diffractive production of 
$N^*_{1440}$- and  $N^*_{1680}$-resonances make also sizeable contribution into 
the reaction $n~p~\rightarrow~p~p~\pi^-$. Therefore it is necessary to take into
account such processes for the reaction $n p \rightarrow n p \pi^+ \pi^-$ that 
are described by the diagrams similar to diagrams in Fig.\ref{Fig7} with 
the replacement of $\pi$-meson exchange by the exchange of vacuum pole (pomeron)
  The matrix element for the diagrams of pomeron exchange is written in 
following form:
\begin{equation}
\label{Pom13}
  T_{N^*}(s,t)=i8 \pi s_0 \; g_{N}^P(t) \; F_{13} \; T_{\pi N \rightarrow  
  N^* \rightarrow \pi \pi N}
\end{equation}

 where $g_N^P(t) = g_N^P(0) \; e^{-R_N^2|t|}$ - vertex function,\\
\hspace*{1.3cm} $\alpha_P(t) = \alpha_P(0)+\alpha_P' \; t$ Regge trajectory of 
pomeron.\\
The values $g_N^P(0)$, $R_N^2$, $\alpha_P(0)$ and $\alpha_P'$ wre taken 
from~\cite{Nikitin}.

   The results of the description of the reaction $n~p~\rightarrow~p~p~\pi^-$ 
by diagrams~(\ref{Fig5}),~(\ref{Fig7}) and pomeron exchange at 
$P_0=5.2$ GeV/c are presented in Fig.~\ref{Fig8}:
\begin{figure}[h]
\hspace{0.5cm}
\includegraphics[width=1.0\textwidth]{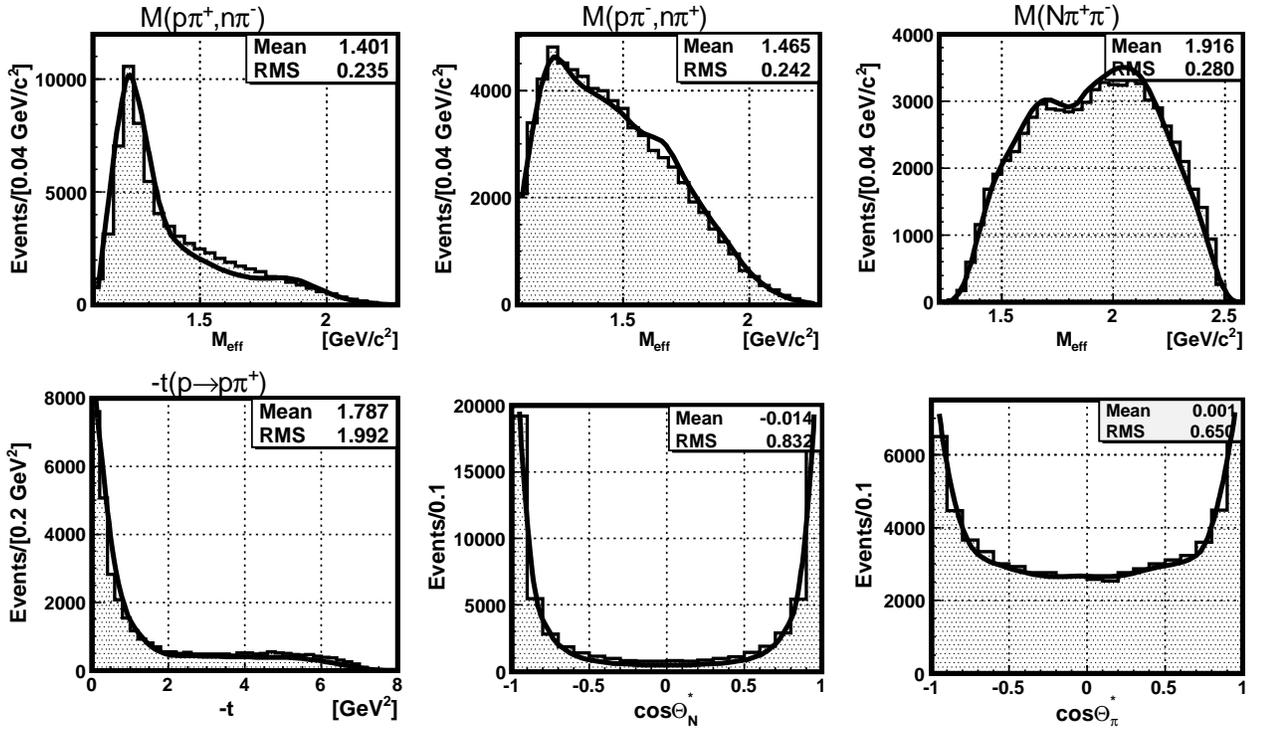}
\caption{Distributions for the reaction $np \rightarrow np \pi^+ \pi^-$ at 
 $P_0$=5.20 GeV/c}
\label{Fig8}
\end{figure}

  One can see a good agreement between the experimental distributions and 
theoretical calculations.

\newpage
\section{The reaction $np \rightarrow np \pi^+ \pi^-$ at $P_0$ $<$ 3 GeV/c}
  The study of effective mass spectra of $np$ - combinations at $P_0$=1.73 and 
2.23 GeV/c 
(Fig.9) shows the clear peack close the threshold ($M_{np}= m_n+m_p$) that can 
not be described within the framework of OPER-model using the diagrams from 
Fig.\ref{Fig5} and Fig.\ref{Fig7}.
\begin{figure}[h]
\includegraphics[width=0.9\textwidth]{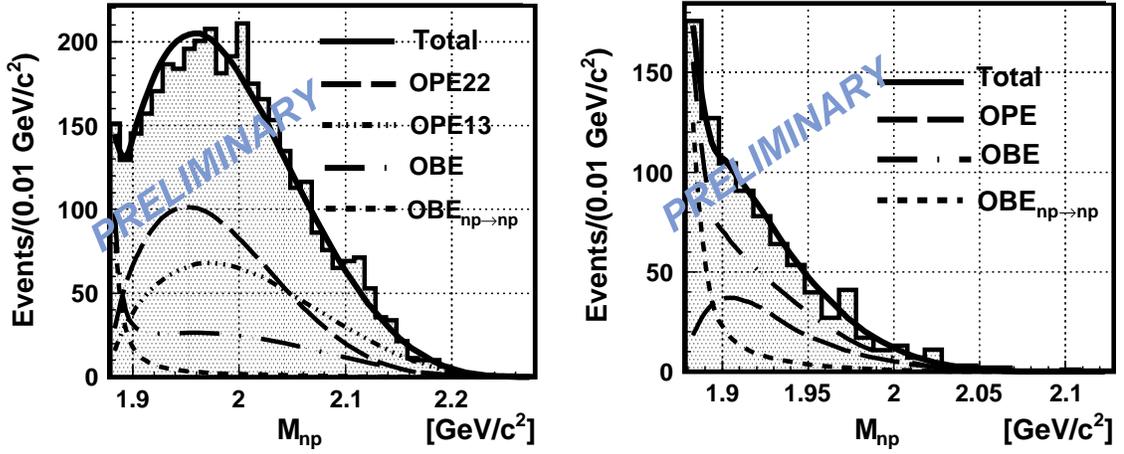}
\caption{The distributions of $M_{np}$ for treaction $np \rightarrow np \pi^+ 
 \pi^-$ at $P_0$ = 2.23 GeV/c (left) and 1.73 GeV/c (right).}
\label{fig:Fig9}
\end{figure}

  The model of Regge poles with baryon exchange and nonlinear trajectories, 
suggested in~\cite{OBE} was used to describe these features. The following 
diagrams of one baryon exchange (OBE) were taken into account  within the
framework of this model:
\begin{figure}[h]
\includegraphics[width=0.85\textwidth]{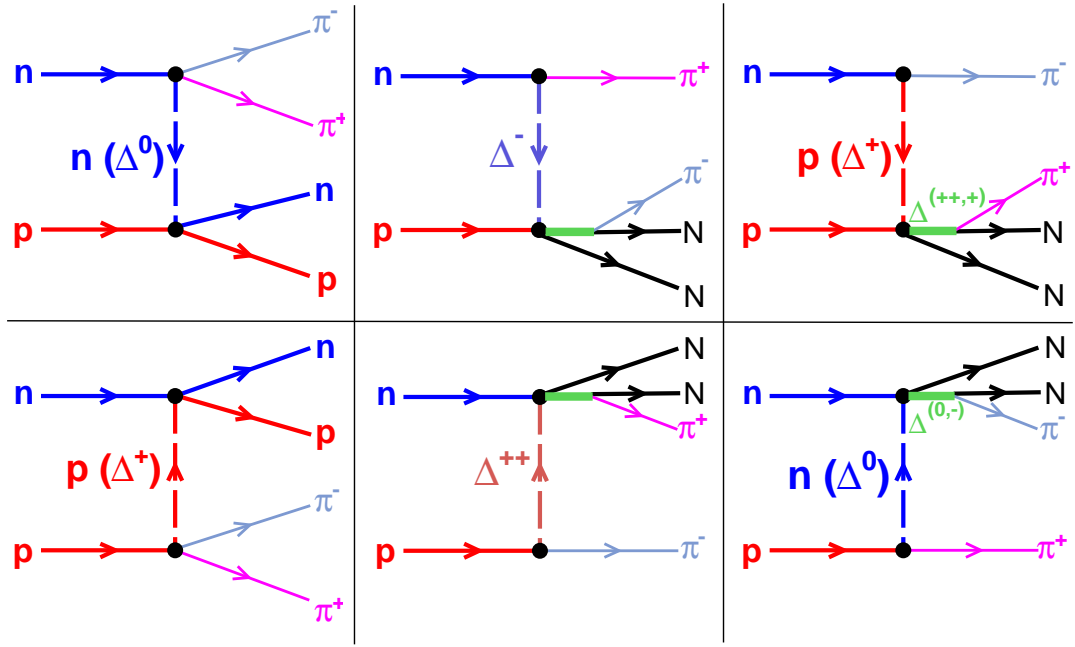}
\caption{OBE diagrams for the reaction $np \rightarrow np \pi^+ \pi^-$}
\label{fig:Fig10}
\end{figure}

   The vertex function of elastic $np \rightarrow np$ scattering was calculated 
using the data from~\cite{NNel}. 

   The vertex functions of $\Delta N \rightarrow n p$, $N N \rightarrow \Delta 
N$ and $\Delta N \rightarrow \Delta N$ scattering were calculated corresponding 
to~\cite{NNND}.
In result  one can get the good description of the experimental distribution 
from the reaction $np \rightarrow np \pi^+ \pi^-$ at $P_0$ = 1.73 and 2.23 GeV/c 
(Fig.9 and Fig.11).
\vspace{-0.1cm}
\begin{figure}[h]
\includegraphics[width=0.9\textwidth]{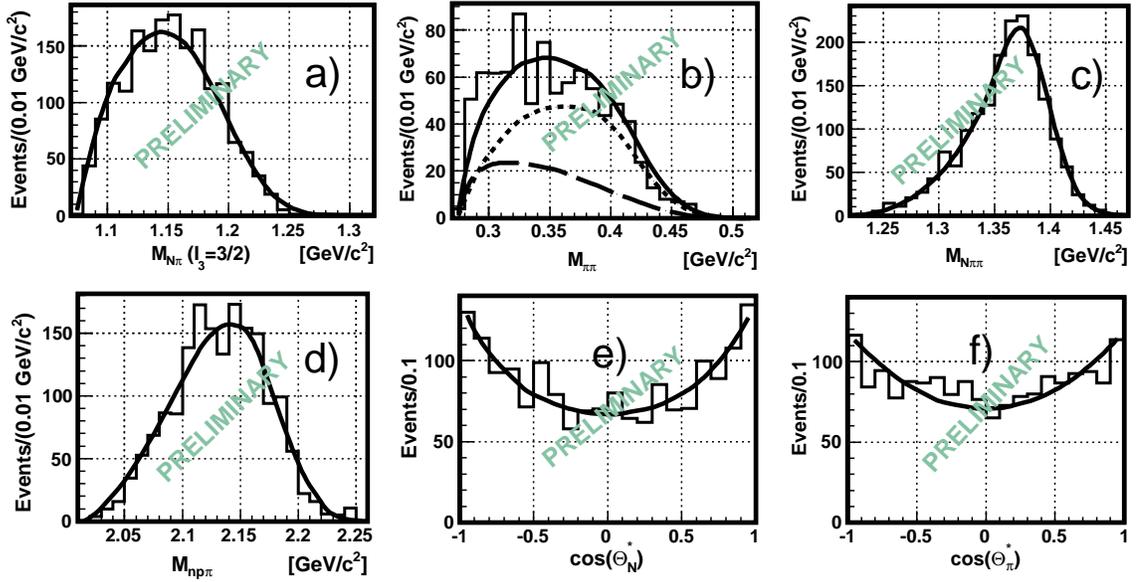}
\caption{Distributions for the reaction $np \rightarrow np \pi^+ \pi^-$ at 
 $P_0$=1.73 GeV/c}
\label{fig:Fig11}
\end{figure}

\newpage
\section{OPER model and other reactions}

 The other reactions of np interactions are scheduled to study by means of 
OPER~model:\\
\hspace*{1.5cm} $n p \rightarrow p p \pi^-$ 
\hspace*{1.55cm} vertex functions 1$\times$2\\
\hspace*{1.5cm} $n p \rightarrow p p \pi^- \pi^0$ 
\hspace*{1.15cm} vertex functions 2$\times$2 and 1$\times$3\\
\hspace*{1.5cm} $n p \rightarrow p p \pi^+ \pi^- \pi^-$ 
\hspace*{0.55cm} vertex functions 2$\times$3\\ 
\hspace*{1.5cm} $n p \rightarrow p p \pi^+ \pi^- \pi^- \pi^0$ 
\hspace*{0.15cm} vertex functions 3$\times$3\\  
\hspace*{1.5cm} $n p \rightarrow n p \pi^+ \pi^- \pi^+ \pi^-$ 
\hspace*{0.05cm} vertex functions 3$\times$3\\

\begin{figure}[h]
\includegraphics[width=1.0\textwidth]{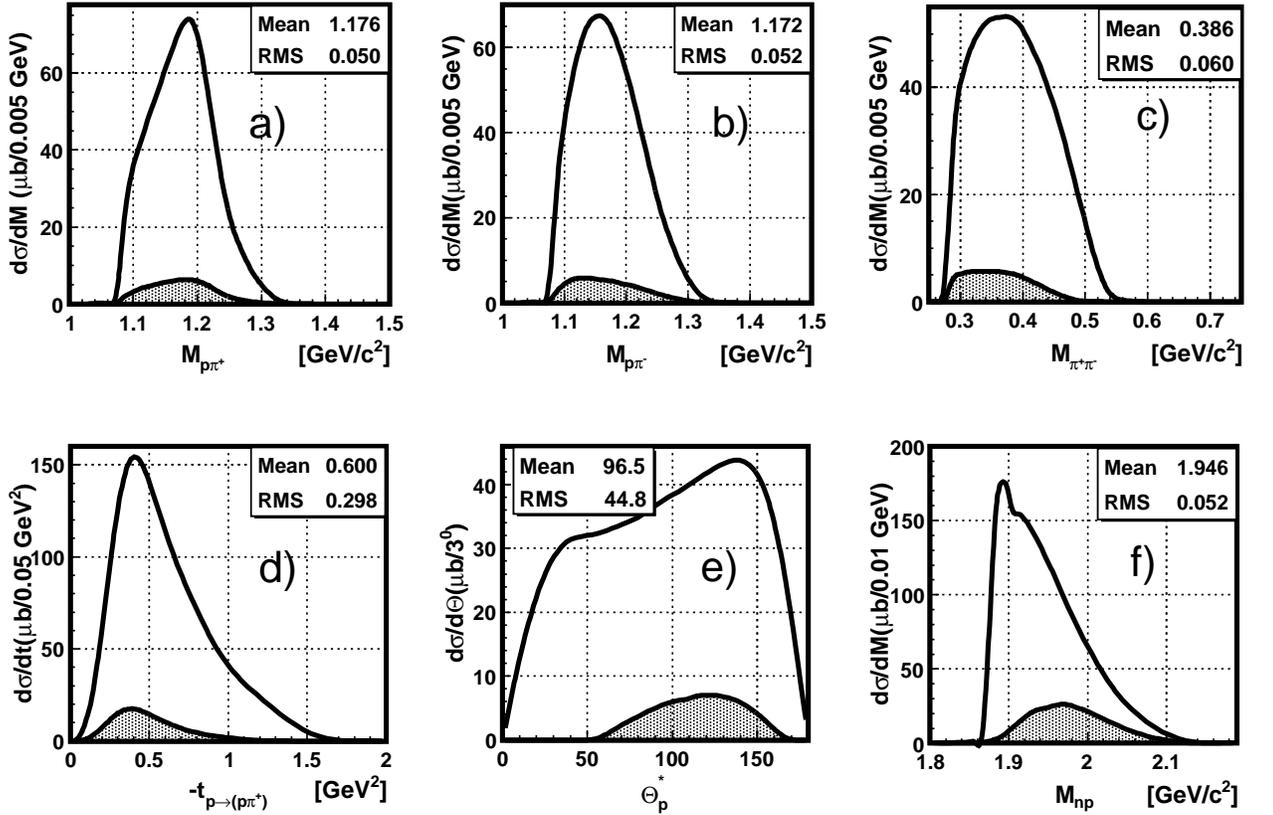}
\caption{Disdributions for the reaction $np \rightarrow np \pi^+ \pi^-$ at 
$T_{kin}$=1.25 GeV,
 calculated for HADES set-up. Dashed area - results of HADES acceptance.}
\label{Fig12}
\end{figure}

   Similar reactions of $pp$, $\bar p p$ and $\pi N$ interactions also can be 
described by OPER~model.
The following reactions were simulated for HADES experiment~\cite{HADES}:\\
\hspace*{1.5cm}  $pp \rightarrow pp \pi^+ \pi^-$ at $T_{kin}$=3.5 GeV\\
\hspace*{1.5cm}  $np \rightarrow np \pi^+ \pi^-$ at $T_{kin}$=1.25 GeV 
(see Fig.\ref{Fig12})\\
\hspace*{1.5cm}  $np \rightarrow np e^+ e^-$     at $T_{kin}$=1.25 GeV with vertex 
 function of $\gamma N \rightarrow N e+ e^-$.\\

   Since the $\pi N \rightarrow \pi N$  and $\pi N \rightarrow \pi \pi N$ vertex 
functions  are taken in helicity representation it seems to be perspective to use 
OPER~model for description of the reaction with polarized particles.

\section{Conclusion}

  Reaction $np \rightarrow np \pi^+ \pi^-$ is characterized by the plentiful 
production of the $\Delta$ resonance and the large peripherality of the secondary 
particles. The experimental data are successfully described by the further development
of OPER~model.\\

 However  at  $P_0$ $<$ 3 GeV/c it is necessary to take into account another mechanism
 of the reaction (such as OBE).\\

  OPER~model permits to describe another $N(\bar N)-N$ reactions with the production 
of some $\pi$-mesons. The further development of OPER-model  can be very promising 
to describe the production of $e^+ e^-$-pairs in hadronic interactions.\\

  OPER~model can be used as an effective tool to simulate various reactions of 
hadronic interactions.

\newpage


\end{document}